\documentclass[twocolumn,aps,pre,showpacs,amsmath,amssymb,amsfonts,floatfix,superscriptaddress]{revtex4}
\usepackage{graphicx}% Include figure files
\usepackage{dcolumn}% Align table columns on decimal point
\usepackage{bm}% bold math
\usepackage[hypertex]{hyperref}
\usepackage{latexsym}
\usepackage{float}
\usepackage{supertabular}
\usepackage{longtable}

\begin{document}
\title{Understanding Baseball Team Standings and Streaks}

\author{C. Sire} %\email{Clement.Sire@irsamc.ups-tlse.fr}
\affiliation{Laboratoire de Physique Th\'eorique - IRSAMC, CNRS,
Universit\'e Paul Sabatier, 31062 Toulouse, France}
\author{S.~Redner} %\email{redner@bu.edu}
\affiliation{Laboratoire de Physique Th\'eorique - IRSAMC, CNRS, Universit\'e Paul
  Sabatier, 31062 Toulouse, France}
\affiliation{Center for Polymer Studies and
  Department of Physics, Boston University, Boston, Massachusetts 02215, USA}

\begin{abstract}

  Can one understand the statistics of wins and losses of baseball teams?
  Are their consecutive-game winning and losing streaks self-reinforcing or
  can they be described statistically?  We apply the Bradley-Terry model,
  which incorporates the heterogeneity of team strengths in a minimalist way,
  to answer these questions.  Excellent agreement is found between the
  predictions of the Bradley-Terry model and the rank dependence of the
  average number team wins and losses in major-league baseball over the past
  century when the distribution of team strengths is taken to be uniformly
  distributed over a finite range.  Using this uniform strength distribution,
  we also find very good agreement between model predictions and the observed
  distribution of consecutive-game team winning and losing streaks over the
  last half-century; however, the agreement is less good for the previous
  half-century.  The behavior of the last half-century supports the
  hypothesis that long streaks are primarily statistical in origin with
  little self-reinforcing component.  The data further show that the past
  half-century of baseball has been more competitive than the preceding
  half-century.
\end{abstract}

\pacs{89.75.-k, 02.50.Cw}

\maketitle

\section{Introduction}

The physics of systems involving large numbers of interacting agents is
currently a thriving field of research \cite{complex}.  One of its many
appeals lies in the opportunity it offers to apply precise methods and tools
of physics to the realm of ``soft'' science.  In this respect, biological,
economic, and a large variety of human systems present many examples of
competitive dynamics that can be studied qualitatively or even quantitatively
by statistical physics.  Among them, sports competitions are particularly
appealing because of the large amount of data available, their popularity,
and the fact that they constitute almost perfectly \emph{isolated} systems.
Indeed, most systems considered in econophysics \cite{bouchaud} or
evolutionary biology \cite{krug} are strongly affected by external and often
unpredictable factors.  For instance, a financial model cannot predict the
occurrence of wars or natural disasters which dramatically affect financial
markets, nor can it include the effect of many other important external
parameters (China's GDP growth, German exports, Google's profit\ldots).  On
the other hand, sport leagues (soccer \cite{foot}, baseball \cite{chance},
football \cite{PN}\ldots) or tournaments (basketball \cite{basket,BH}, poker
\cite{poker}\ldots) are basically isolated systems that are much less
sensitive to external influences.  Hence, despite their intrinsic human
nature, which actually contribute to their appeal, competitive sports are
particularly suited to quantitative theoretical modeling.  In this spirit,
this work is focused on basic statistical features of game outcomes in
Major-League baseball.

In Major-League baseball and indeed in any competitive sport, the main
observable is the outcome of a single game --- who wins and who loses. Then
at the end of a season, the win/loss record of each team is fundamental.  As
statistical physicists, we are not concerned with the fates of individual
teams, but rather with the average win/loss record of the $1^{\rm st}$,
$2^{\rm nd}$, $3^{\rm rd}$, {\it etc.}  teams, as well as the statistical
properties of winning and losing streaks.  We concentrate on major-league
baseball to illustrate statistical properties of game outcomes because of the
large amount of available data \cite{data} and the near constancy of the game
rules during the so-called ``modern era'' that began in 1901.

For non-US readers or for non-baseball fans, during the modern era of
major-league baseball, teams have been divided into the nearly-independent
American and National leagues \cite{inter}.  At the end of each season a
champion of the American and National leagues is determined (by the best team
in each league prior to 1961 and by league playoffs subsequently) that play
in the World Series to determine the champion.  As the data will reveal,
it is also useful to separate the 1901--1960 early modern era, with a
154-game season and 16 teams, and the 1961--2005 expansion era, with a
162-game season in which the number of teams expanded in stages to its
current value of 30, to highlight systematic differences between these two
periods.  Our data is based on the 163674 regular-season games that have
occurred between 1901 and the end of the 2005 season (72741 between 1901--60
and 90933 between 1961--2005).

While the record of each team can change significantly from year to year, we
find that the {\em time average} win/loss record of the $r^{\rm th}$-ranked
team as a function of rank $r$ is strikingly regular.  One of our goals is to
understand the rank dependence of this win fraction.  An important outcome of
our study is that the Bradley-Terry (BT) competition model \cite{Z29,BT52}
provides an excellent account of the team win/loss records.  This agreement
between the data and theory is predicated on using a specific form for the
distribution of team strengths.  We will argue that the best match to the
data is achieved by using a uniform distribution of teams strengths in each
season.

Another goal of this work is to understand the statistical features of
consecutive-game team winning and losing streaks.  The existence of long
streaks of all types of exceptional achievement in baseball, as well as in
most competitive sports, have been well documented \cite{streak-data} and
continue to be the source of analysis and debate among sports fans.  For long
consecutive-game team winning and team losing streaks, an often-invoked theme
is the notion of reinforcement---a team that is ``on a roll'' is more likely
to continue winning, and vice versa for a slumping team on a losing streak.
The question of whether streaks are purely statistical or self reinforcing
continues to be vigorously debated \cite{streaks}.  Using the BT model and
our inferred uniform distribution of team strengths, we compute the streak
length distribution.  We find that the theoretical prediction agrees
extremely well with the streak data during 1961--2005.  However, there is a
slight discrepancy between theory and the tail of the streak distribution
during 1901--60, suggesting that non-statistical effects may have played a
role during this early period.

As a byproduct of our study, we find clear evidence that baseball has been
more competitive during 1961--2005 than during 1901--60 and feature that has
been found previously \cite{BVR}.  The manifestation of this increased
competitiveness is that the range of team records and the length of streaks
was narrower during the latter period.  This observation fits with the
general principle \cite{gould} that outliers become progressively rarer in a
highly competitive environment.  Consequently, extremes of achievement become
less and less likely to occur.

\section{Statistics of the Win Fraction}

\subsection{Bradley-Terry Model}

Our starting point to account for the win/loss records of all baseball teams
is the BT model \cite{Z29,BT52} that incorporates the heterogeneity in team
strengths in a natural and simple manner.  We assume that each team has an
intrinsic strength $x_i$ that is fixed for each season.  The probability that
a team of strength $x_i$ wins when it plays a team of strength $x_j$ is
simply
\begin{equation}
\label{pij}
p_{ij} = \frac{x_i}{x_i+x_j}~.
\end{equation}
Thus the winning probability depends continuously on the strengths of the two
competing teams \cite{contrast}.  When two equal-strength teams play, each
team has a 50\% probability to win, while if one team is much stronger, then
its winning probability approaches 1.

The form of the winning probability of Eq.~\eqref{pij} is quite general.
Indeed, we can replace the team strength $x_i$ by any monotonic function
$f(x_i)$.  The only indispensable attribute is the ordering of the team
strengths.  Thus the notion of strength is coupled to the assumed form of the
winning probability.  If we make a hypothesis about one of these quantities,
then the other is no longer a variable that we are free to choose, but an
outcome of the model.  In our analysis, we adopt the form of the winning
probability in Eq.~\eqref{pij} because of its simplicity.  Then the only
relevant unknown quantity is the probability distribution of the $x_i$'s. As
we shall see in the next section, this distribution of team strengths can
then be inferred from the season-end win/loss records of the teams, and a
good fit to the data is obtained when assuming a uniform distribution of team
strengths.  Because only the ratio of team strengths is relevant in
Eq.~\eqref{pij}, we therefore take team strengths to be uniformly distributed
in the range $[x_{\rm min},1]$, with $0\leq x_{\rm min} \leq 1$.  Thus the
only model parameter is the value of $x_{\rm min}$.

\begin{figure}[ht]
\centerline{\includegraphics*[width=0.45\textwidth]{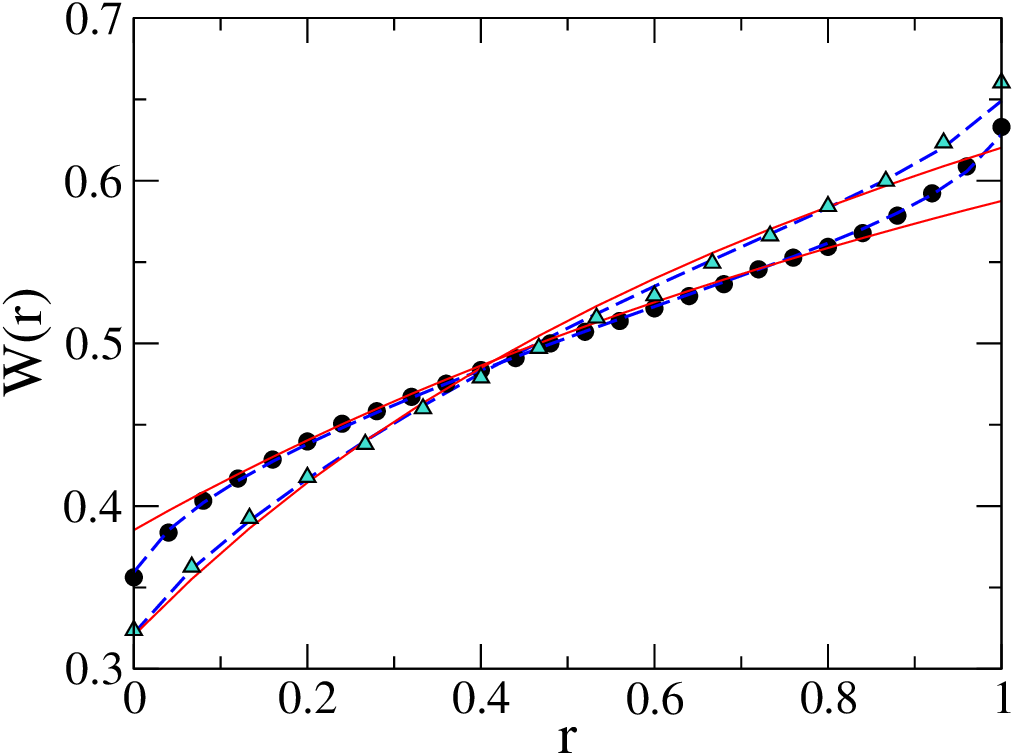}}
\caption{Average win fraction $W(r)$ versus scaled rank $r$ for 1901--60
  ($\bigtriangleup$) and 1961--2005 ($\circ$).  For these periods, the dashed
  lines are simulation results for the BT model with $x_{\rm min}=0.278$ and
  $0.435$ respectively.  The solid curves represent Eq.~\eqref{W},
  corresponding to simulations for an infinitely long season and an infinite
  number of teams.}
  \label{w-frac}
\end{figure}

For uniformly distributed team strengths $\{x_j\}$ that lie in $[x_{\rm
  min},1]$, the average winning fraction for a team of strength $x$ that
plays a large number of games $N$, with equal frequencies against each
opponent is
\begin{eqnarray}
\label{W}
W(x) &=& \frac{1}{N}\sum_{j=1}^N \frac{x}{x+x_j}\nonumber \\
&\to& \frac{x}{1-x_{\rm min}} \int_{x_{\rm min}}^1 \frac{dy}{x+y} \nonumber \\
&=&\frac{x}{1-x_{\rm min}}\ln\left(\frac{x+1}{x+x_{\rm min}}\right)~,
\end{eqnarray}
where we assume $N\to\infty$ in the second line.  We then transform from
strength $x$ to scaled rank $r$ by $x=x_{\rm min} +(1-x_{\rm min})r$, with
$r=0,1$ corresponding to the weakest and strongest team, respectively
(Fig.~\ref{w-frac}).  This result for the win fraction is one of our primary
results.

\begin{figure}[ht]
\includegraphics*[width=0.45\textwidth]{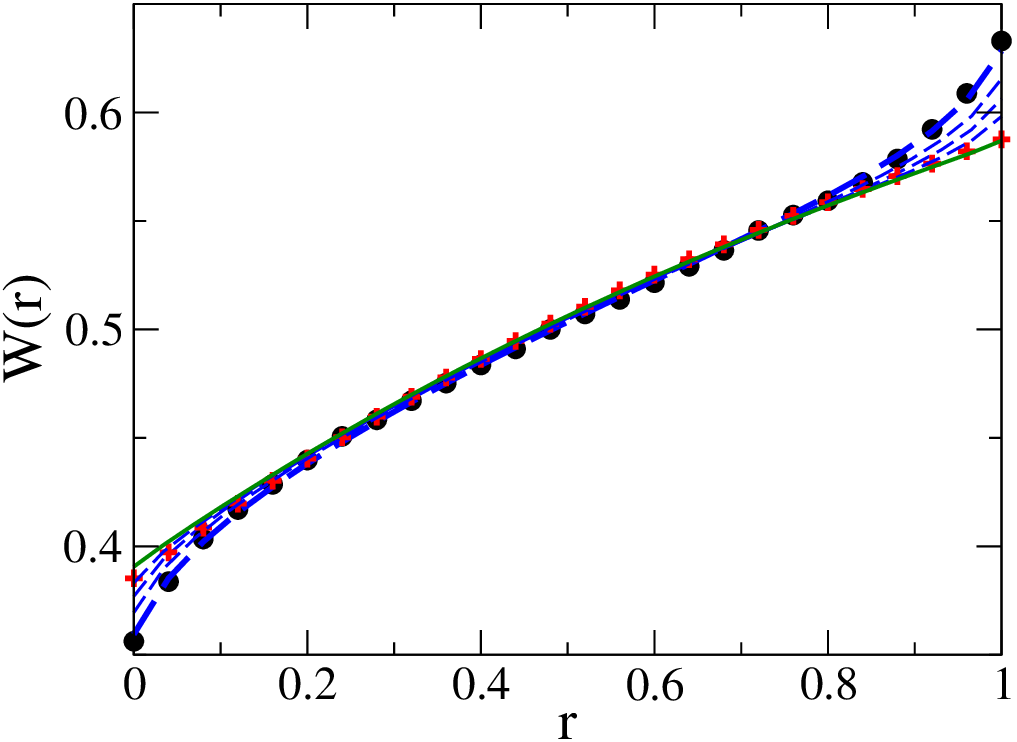}
\caption{Convergence of $W(r)$ versus scaled rank $r$ as a function of season
  length for 1961--2005, using $x_{\rm min}=0.435$ and 30 teams.  The circles
  and the thick dashed curve are the baseball data and the corresponding BT
  model data for a $n=162$ game season.  The thin dashed lines are model data
  for a season of $n=300, 500$, and 1000 games averaged over 100000 seasons.
  The full line corresponds to the model for an infinitely long season with
  30 teams. Finally, the $+$ symbols give the result of Eq.~\eqref{W}, which
  corresponds to an infinite-length season and an infinite number of teams.}
  \label{w-conv}
\end{figure}

To check the prediction of Eq.~\eqref{W}, we start with a value of $x_{\rm
  min}$ and simulate $10^4$ periods of a model baseball league that consists
of: (i) 16 teams that play 60 seasons of 154 games (corresponding to
1901--60) and (ii) 30 teams that play 45 seasons of 162 games (1961--2005),
with uniformly distributed strengths in $[x_{\rm min},1]$ for both cases, but
with different values of $x_{\rm min}$.  Using the winning probability
$p_{ij}$ of Eq.~\eqref{pij}, we then compute the average win fraction $W(r)$
of each team as function of its scaled rank $r$.  We then incrementally
update the value of $x_{\rm min}$ to minimize the difference between the
simulated values of $W(r)$ with those from game win/loss data.  Nearly the
same results are found if each team plays every opponent with equal
probability or equally often, as long as the number of teams and number of
games is not unrealistically small.  The BT model, with each team playing
each opponent with the same probability, gives very good fits to the data by
choosing $x_{\rm min}=0.278$ for the period 1901--60, and $x_{\rm min}=0.435$
for 1961--2005 (Fig.~\ref{w-frac}).  If the actual game frequencies in each
season are used to determine opponents, $x_{\rm min}$ changes slightly---to
0.289 for 1901--60---but remains unchanged for 1961--2005.

Despite the fact that the number of teams has increased from 16 to 30 since
in 1961, the range of win fractions is larger in the early era (0.32--0.67)
than in the expansion era (0.36--0.63), a feature that indicates that
baseball has become more competitive.  This observation accords with the
notion that the pressure of continuous competition, as in baseball, gradually
diminishes the likelihood of outliers \cite{gould}.  Given the crudeness of
the model and real features that we have ignored, such as home-field
advantage (approximately 53\% for the past century and slowly decreasing with
time), imbalanced playing schedules, and in-season personnel changes due to
trades and player injuries, the agreement between the data and simulations of
the BT model is satisfying.

It is worth noting in Fig.~\ref{w-frac} is that the win fraction data and
the corresponding numerical results from simulations of the BT model deviate
from the theoretical prediction given in Eq.~\eqref{W} when $r\to 0$ and
$r\to 1$.  This discrepancy is simply a finite-season effect.  As shown in
Fig.~\ref{w-conv}, when we simulate the BT model for progressively longer
seasons, the win/loss data gradually converges to the prediction of
Eq.~\eqref{W}.

The present model not only reproduces the average win record $W(r)$ over a
given period, but it also correctly explains the season-to-season fluctuation
$\sigma^2(r)$ of the win fraction defined as
\begin{eqnarray}
\label{sig}
\sigma^2(r) &\equiv& \frac{1}{Y}\sum_{j=1}^Y (W(r)-W_j(r))^2,
\end{eqnarray}
where $W_j(r)$ is the winning fraction of the $r^{\rm th}$-ranked team during
the $j^{\rm th}$ season and
\begin{eqnarray}
W(r)&= &\frac{1}{Y}\sum_{j=1}^Y W_j(r),\nonumber
\end{eqnarray}
is the average win fraction of the $r^{\rm th}$-ranked team and $Y$ is the
number of years in the period.  These fluctuations are the largest for
extremal teams (and minimal for average teams).  There is also an asymmetry
of $\sigma(r)$ with respect to $r=1/2$.  Our simulations of the BT model with
the optimal $x_{\rm min}$ values that were determined previously by fitting
to the win fraction quantitatively reproduce these two features of
$\sigma(r)$.

\begin{figure}[ht]
\includegraphics*[width=0.45\textwidth]{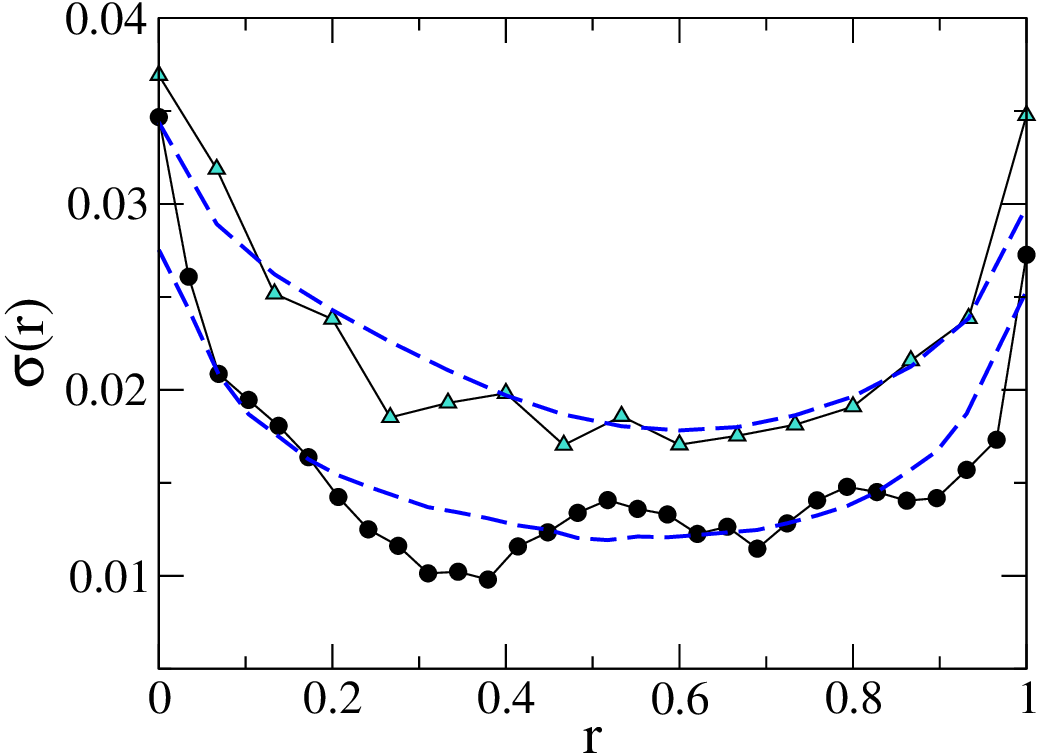}
\caption{Season-to-season fluctuation $\sigma(r)$ for 1901--60
  ($\bigtriangleup$) and for 1961--2005 ($\circ$).  The dashed lines are
  numerical simulations of the BT model for $10^4$ periods with the same
  $x_{\rm min}$ as in Fig.~\ref{w-frac}.}
  \label{sigma}
\end{figure}

In addition to the finite-season effects described above, another basic
consequence of the finiteness of the season is that the intrinsically
strongest team does not necessarily have the best win/loss record.  That is,
the average win fraction $W$ does not necessarily increase with team
strength.  By luck, a strong team can have a poor record or vice versa.  It
is instructive to estimate the number of games $G$ that need to be played to
ensure that the win/loss record properly reflects team strength.  The
difference in the number of wins of two adjacent teams in the standings is
proportional to $G{\times}(1-x_{\rm min})/T$, namely, the number of games times
their strength difference; the latter is proportional to $(1-x_{\rm min})/T$
for a league that consists of $T$ teams.  This systematic contribution to the
difference should significantly exceed random fluctuations, which are of the
order of $\sqrt{G}$.  Thus we require
\begin{equation}
\label{G}
G\gg \left(\frac{T}{1-x_{\rm min}}\right)^2
\end{equation}
for the end-of-season standings to be ordered by team strength.
Fig.~\ref{w-conv} and Fig.~\ref{sigma} illustrate the fact that this effect
is more important for the top-ranked and bottom-ranked teams.  During the
1901--60 period, when major-league baseball consisted of independent American
and National leagues, $T=8$, $G=154$, and $x_{\rm min}\approx 0.3$, so that
the season was just long enough to resolve adjacent teams.  Currently,
however, the season length is insufficient to resolve adjacent teams.  The
natural way to deal with this ambiguity is to expand the number of teams that
qualify for the post-season playoffs, which is what is currently done.

\subsection{Applicability of the Bradley-Terry Model}

Does the BT model with uniform teams strength provide the most appropriate
description of the win/loss data?  We perform several tests to validate this
model.  First, as mentioned in the previous section, the assumption
\eqref{pij} for the winning probability can be recast more generally as
\begin{equation}
\label{gen}
p_{ij}=\frac{f(x_i)}{f(x_i)+f(x_j)}~,
\end{equation}
so that an arbitrary $X_i=f(x_i)$ reduces to the original winning probability
in Eq.~\eqref{pij}.  Hence the crucial model assumption is the separability
of the winning probability.  In particular, the BT model assumes that
${p_{ij}}/{p_{ji}}=p_{ij}/(1-p_{ij})$ is {\em only} a function of
characteristics of team $i$, divided by characteristics of team $j$.  One
consequence of this separability is the ``detailed-balance'' relation
\begin{equation}
\label{db1}
\frac{p_{ik}}{1-p_{ik}}{\times} \frac{p_{kj}}{1-p_{kj}} = \frac{p_{ij}}{1-p_{ij}},
\end{equation}
for any triplet of teams.  This relation quantifies the obvious fact that if
team $A$ likely beats $B$, and $B$ likely beats $C$, then $A$ is likely to
beat $C$.  Since we do not know the actual $p_{ij}$ in a given baseball
season, we instead consider
\begin{equation}
\label{z}
z_{ij}=\frac{W_{ij}}{G_{ij}-W_{ij}},
\end{equation}
where $W_{ij}$ is the number of wins of team $i$ against $j$, and $G_{ij}$ is
the number of game they played against each other in a given season.  If
seasons were infinitely long, then $z_{ij}\to p_{ij}/(1-p_{ij})$, and hence
\begin{equation}
\label{dbz}
z_{ik}{\times} z_{kj}=z_{ij}.
\end{equation}

\begin{figure}[ht]
\includegraphics*[width=0.44\textwidth]{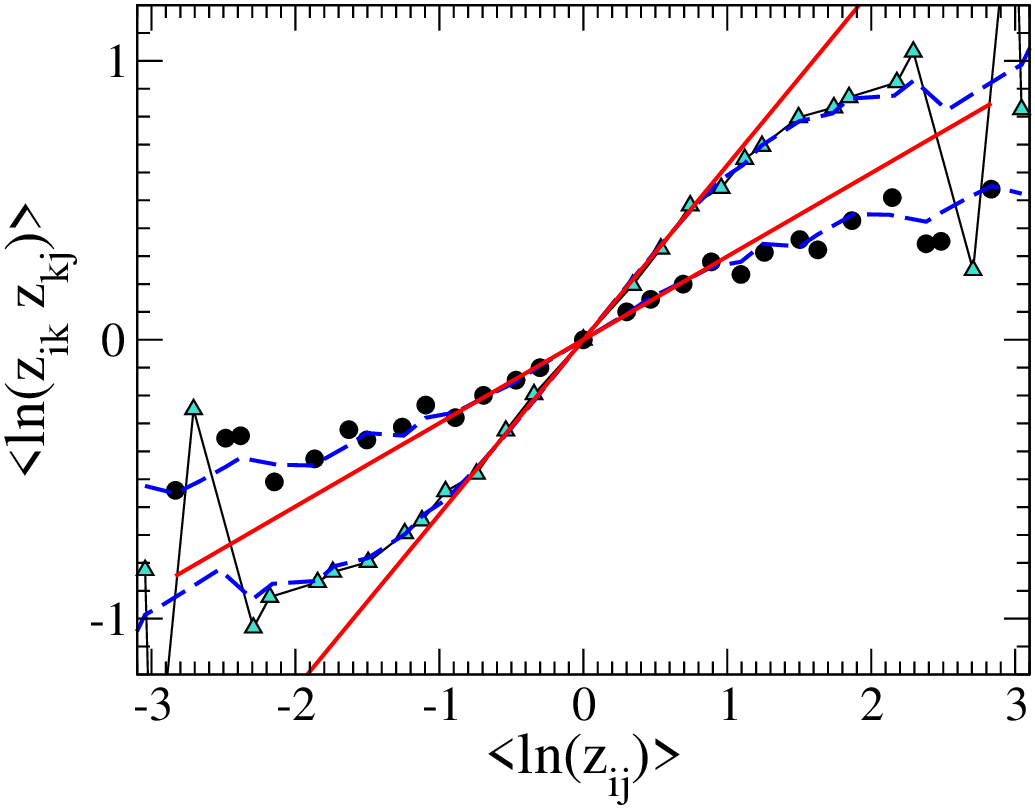}
\caption{Comparison of the detailed balanced relation Eq.~\eqref{dbz} for
  baseball data to the results of the BT model over $10^4$ periods (dashed
  lines), where each period corresponds to the results of all baseball games
  during either 1901--60 (triangles) or 1960--2005 (circles).  The $x_{\rm
    min}$ values are the same as in Fig.~\ref{w-frac}.  The straight lines
  are guides for the eye, with slope 0.63 for the data for 1901--60 and 0.30
  for 1961--2005.}
  \label{DB}
\end{figure}

\begin{figure}[ht]
\includegraphics*[width=0.46\textwidth]{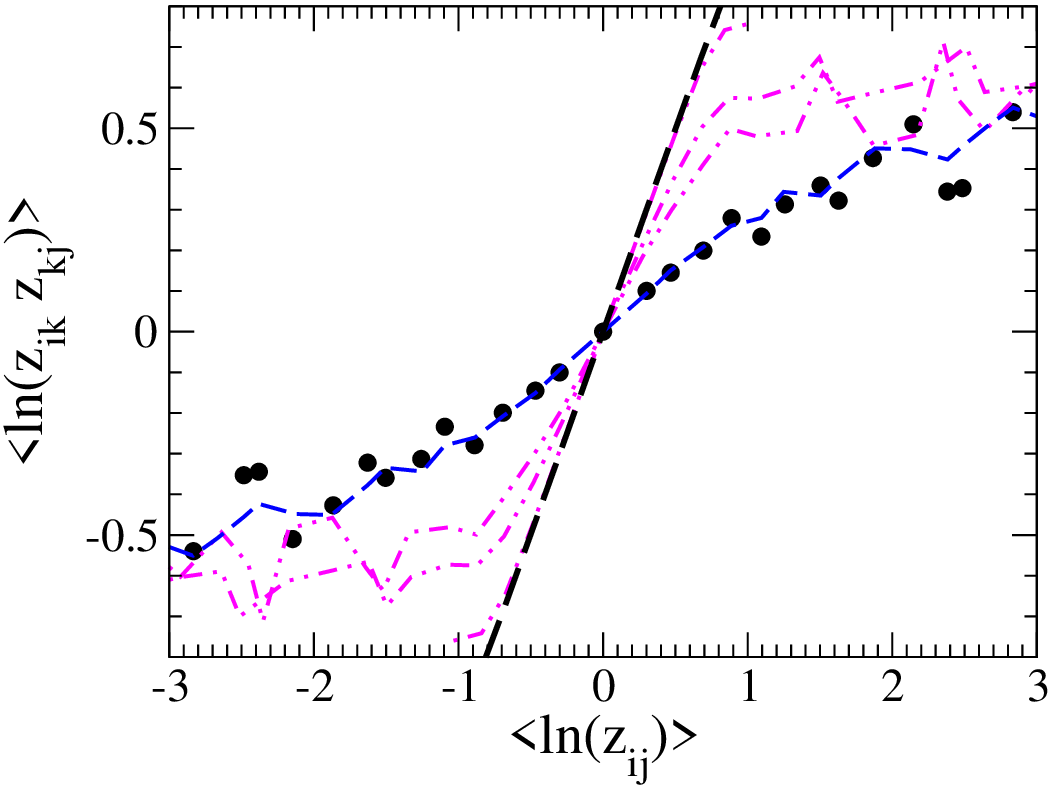}
\caption{Dependence of $\langle \ln(z_{ik}z_{kj})\rangle$ $vs$ $\langle
  \ln(z_{ij})\rangle$ on season length for the 1961--2005 period.  All
  $G_{ij}$'s are multiplied by $M=5$, 10, 100 (steepening dot-dashed lines).
  The thick dashed line corresponds to $M=10^4$ and is indistinguishable from
  a linear dependence with unit slope.}
  \label{DBII}
\end{figure}

To test the detailed balance relation Eq.~\eqref{dbz}, we plot $\langle
\ln(z_{ik}z_{kj})\rangle$ as a function of $\langle \ln(z_{ij})\rangle$ from
game data, averaged over all team triplets $(i,j,k)$ and all seasons in a
given period (Fig.~\ref{DB}).  We discard events for which $W_{ij}=G_{ij}$ or
$W_{ij}=0$ (team $i$ won or lost all games against team $j$).  Our
simulations of the BT model over $10^4$ realizations of the 1901--60 and
1961--2005 periods with the same $G_{ij}$ as in actual baseball seasons and
with the optimal values of $x_{\rm min}$ for each period are in excellent
agreement with the game data.  Although $z_{ik}z_{kj}$ in the figure has a
sublinear dependence of $z_{ij}$ (slope much less than 1 in Fig.~\ref{DB}),
the slope progressively increases and ultimately approaches the expected
linear relation between $z_{ik}z_{kj}$ and $z_{ij}$ as the season length is
increased (Fig.~\ref{DBII}).  We implement an increased season length by
multiplying all the $G_{ij}$ by the same factor $M$.  Notice also that
$\langle \ln(z_{ik}z_{kj})\rangle$ versus $\langle \ln(z_{ij})\rangle$ for
the 1901--60 period has a larger slope than for 1961-2005 because the
$G_{ij}$'s are larger in the former period ($G_{ij}=22$) than in the latter
($G_{ij}$ in the range 5--19).

This study of game outcomes among triplets of teams provides a detailed and
non-trivial validation for the BT form Eq.~\eqref{W} for the winning
probability. As a byproduct, we learn that cyclic game outcomes, in which
team $A$ beats $B$, $B$ beats $C$, and $C$ beats $A$, are unlikely to occur.

\subsection{Distribution of Team Strengths}

Thus far, we have used a uniform distribution of team strengths to derive the
average win fraction for the BT model.  We now determine the most likely
strength distribution by searching for the distribution that gives the best
fit to the game data for $W(r)$ by minimizing the deviation $\Delta$ between
the data and the simulated form of $W(r)$.  Here the deviation $\Delta$ is
defined as
\begin{equation}
\Delta^2=\frac{\sum_r [W(r)-W(r;\rho)]^2}{\sum_r W(r)^2},\label{error}
\end{equation}
where $W(r;\rho)$ is the winning fraction in simulations of the BT model for
a trial distribution $\rho(x)$ in which the actual game frequencies $G_{ij}$
were used in the simulation, and $W(r)$ is the game data for the winning
fraction.

\begin{figure}[ht]
\includegraphics*[width=0.45\textwidth]{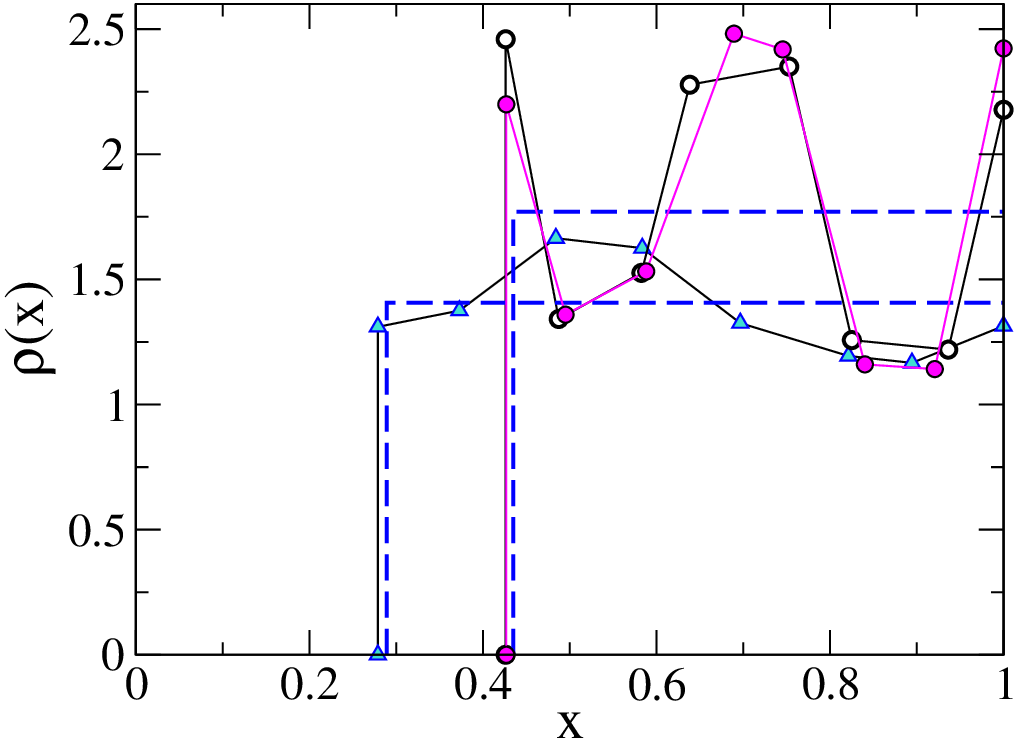}
\caption{Optimized strength distributions $\rho(x)$ for 1901--60 (triangles) and
  1961--2005 (circles), together with the optimal uniform distributions
  (dashed).  For 1961--2005, we also show the final distributions starting
  from $y_i$'s equally spaced between $y_1=0.1$ and $y_8=1$ with the
  distribution $\rho$: (a) uniform on $[0.1,1]$ (open circles), and (b) a
  symmetric V-shape on $[0.1,1]$ (full circles).}
\label{Distx}
\end{figure}

We assume that the two periods 1901--60 and 1961--2005 are long enough for
$W(r)$ to converge to its average value.  We parameterize the trial strength
distribution as a piecewise linear function of $n$ points, $\{\rho(y_i)\}$,
with $y_i\in[0,1]$ and $y_n\equiv 1$.  We then perform Monte Carlo (MC)
simulations, in which we update the $y_i$ and $\rho_i=\rho(y_i)$ by small
amounts in each step to reduce $\Delta$.  Specifically, at each MC step, we
select one value of $i=1,...,n$, and
\begin{itemize}
\item with probability 1/2 adjust $y_i$ (except $y_n=1$) by $\pm u\,\delta
  y/10$, where $\delta y$ is the spacing between $y_i$ and its nearest
  neighbor, and $u$ is a uniform random number between 0 and 1;
\item with probability 1/2, update $\rho(y_i)$ by $\pm u\,\rho(y_i)/10$.
\end{itemize}
If $\Delta$ decreases as a result of this update, then $y_i$ or $\rho(y_i)$
is set to its new value; otherwise the change in the parameter value is
rejected.  We choose $n=8$, which is large enough to obtain a distribution
with significant features and for which typically 1000--2000 MC steps are
sufficient for convergence.  A larger $n$ greatly increases the number of MC
steps necessary to converge and also increases the risk of being trapped in a
metastable state because the size of the phase space grows exponentially with
$n$.  To check that this algorithm does not get trapped in a metastable
state, we started from several different initial states and found virtually
identical final distributions (Fig.~\ref{Distx}).  The MC-optimized
distribution for each period is remarkably close to uniform, as shown in this
figure.

\begin{figure}[ht]
\includegraphics*[width=0.45\textwidth]{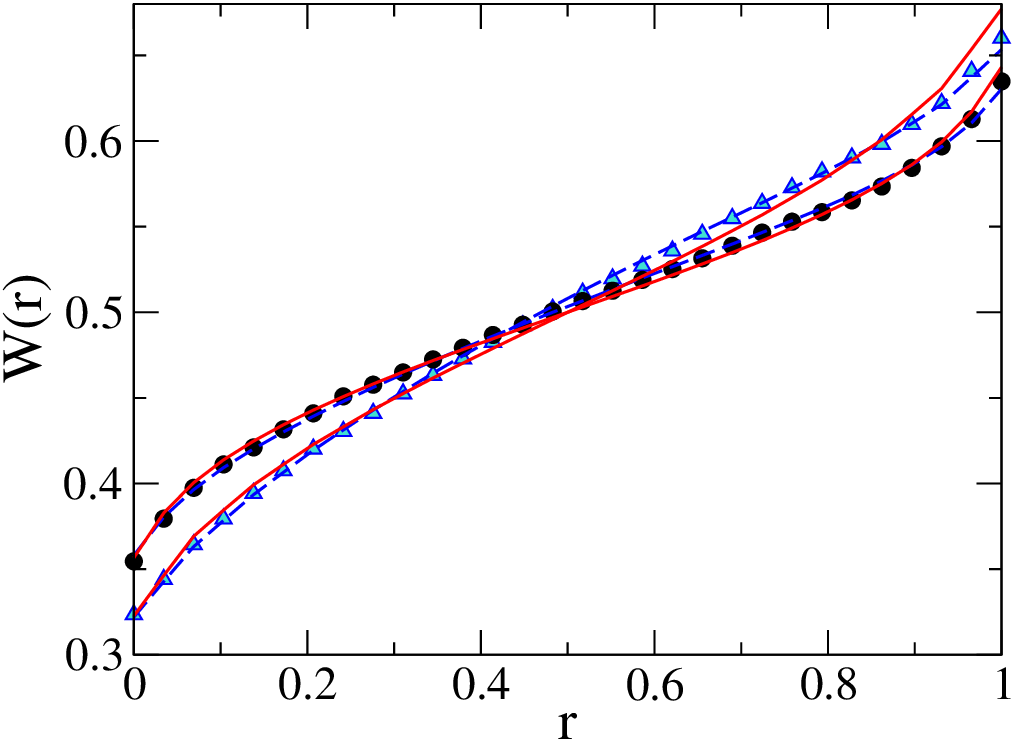}
\caption{Comparison of the winning fraction $W(r)$ extracted from the actual
  baseball data (symbols) to the model with a constant $\rho(x)$ (dashed
  lines), and with the optimal log-normal distribution $\rho(x)$ (full
  lines).}
  \label{fxbestlog}
\end{figure}

Although the optimal distributions are visually not uniform, the small
difference in the relative errors, the closeness of $y_1$ and $x_{\rm min}$,
and the imperceptible difference in the $r$ dependence of $W(r)$ for the
uniform and optimized strength distributions suggests that a uniform team
strength distribution on $[x_{\rm min},1]$ describes the game data quite
well.

For completeness, we also considered the conventionally-used log-normal
distribution of team strengths \cite{JAS93,chance}:
\begin{equation}
\label{LN}
\rho(x)=\frac{1}{\sqrt{2\pi}\kappa x} \exp\left[-\frac{1}{2\kappa^2}\left(
\ln\left(\frac{x}{\bar x}\right)+\frac{\kappa^2}{2}
\right)^2\right].
\end{equation}
With the normalization convention of Eq.~\eqref{LN}, the average team
strength is simply $\bar x$, which can be set to any value due to the
invariance of $p_{ij}$ with respect to the transformation $x\to\lambda x$.
Hence, the only relevant parameter is the width $\kappa$.  Using the same MC
optimization procedure described above, we find that a log-normal {\it
  ansatz} for the strength distribution with optimal parameter $\kappa$ gives
a visually inferior fit of the winning fraction in both periods compared to
the uniform strength distribution, especially for $r$ close to 1 (see
Fig.~\ref{fxbestlog}).  The relative error for the log-normal distribution is
also a factor of 6 and 3 larger, respectively, than for the optimal
distribution in the 1901--60 and 1961--2005 periods.  However, we do
reproduce the feature that the optimal log-normal distribution for 1961--2005
is narrower ($\kappa=0.238$) than that for 1901--60 ($\kappa=0.353$),
indicating again that baseball is more competitive in the second period than
in the first.

\section{Winning and Losing Streak Statistics}

We now turn to the distribution of consecutive-game winning and losing
streaks.  Namely, what are the probabilities $W_n$ and $L_n$ to observe a
string of $n$ consecutive wins or $n$ consecutive losses, respectively?
Because of its emotional appeal, streakiness in a wide variety of sports
continues to be vigorously researched and debated \cite{GVT85,A04,streaks}.
In this section, we argue that independent game outcomes that depend only on
relative team strengths describes the streak data for the period 1961-2005
quite well.  The agreement is not as good for the period 1901-60 and suggests
that non-statistical effects may have played a role in the longest streaks.

Historically, the longest team winning streak (with ties allowed) in
major-league baseball is 26 games, achieved by the 1916 New York Giants in
the National League over a 152-game season \cite{longest}.  The record for a
pure winning streak since 1901 (no ties) is 21 games, set by the Chicago Cubs
in 1935 in a 154-game season, while the American League record is a 20-game
winning streak by the 2002 Oakland Athletics over the now-current 162-game
season.  Conversely, the longest losing streak since 1901 is 23, achieved by
the 1961 Philadelphia Phillies in the National League \cite{worst}, and the
American League losing-streak record is 21 games, set by the Baltimore
Orioles at the start the 1988 season.  For completeness, the list of all
winning and all losing streaks of $\geq$ 15 games is given in the appendix.

\begin{figure}[ht]
\includegraphics*[width=0.45\textwidth]{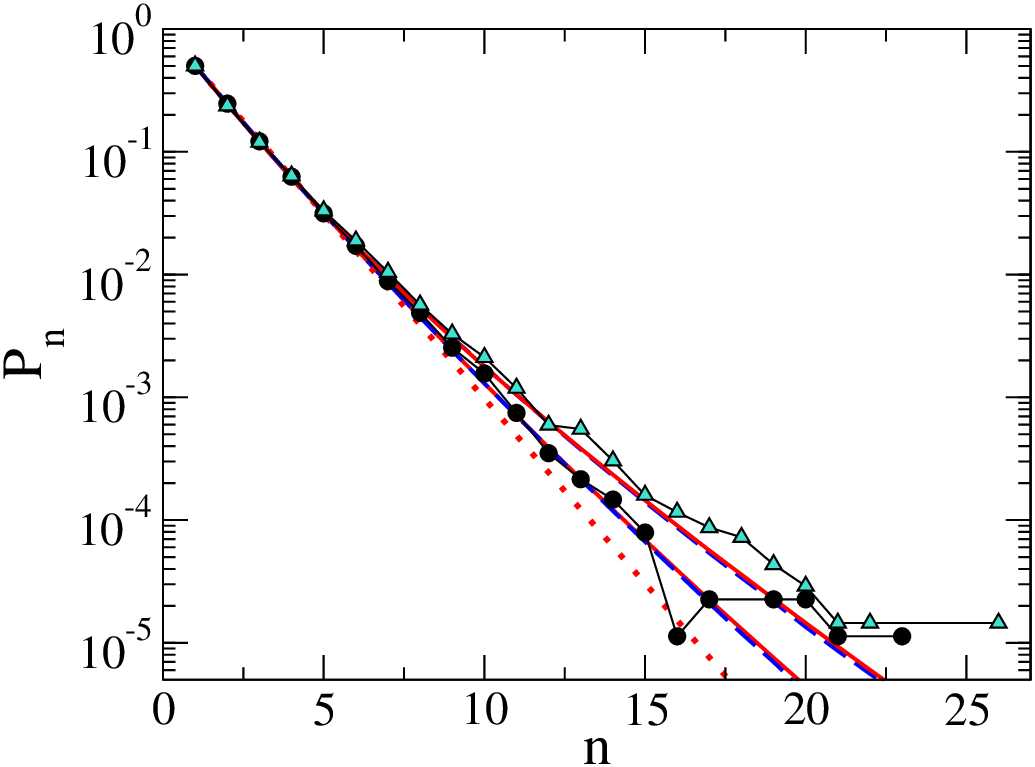}
\caption{Distribution of winning/losing streaks $P_n$ versus $n$ since 1901
  on a semi-logarithmic scale for 1901--60 ($\bigtriangleup$) and 1961--2005
  ($\bullet$).  The dashed curves are the result of simulations with $x_{\rm
    min}=0.278$ and $x_{\rm min} =0.435$ for the two respective periods.  The
  smooth curves are streak data from randomized win/loss records, and the
  dotted curve is $2^{-n}$.}
  \label{streaks-sep}
\end{figure}

Fig.~\ref{streaks-sep} shows the distribution of team winning and losing
streaks in major-league baseball since 1901.  Because these winning and
losing streak distributions are virtually identical for $n\leq 15$, we
consider $P_n=(W_n+L_n)/2$, the probability of a winning {\em or\/} a losing
streak of length $n$ (Fig.~\ref{streaks-sep}).  It is revealing to separate
the streak distributions for 1901--60 and 1961--2005.  Their distinctness is
again consistent with the hypothesis that baseball is becoming more
competitive.  In fact, exceptional streaks were much more likely between
1901--60 than after 1961.  Of the 55 streaks of $\ge 15$ games, 27 occurred
between 1901--30, 13 between 1931--60, and 15 after 1960 \cite{expansion}.

The first point about the streak distributions is that they decay
exponentially with $n$, for large $n$.  This behavior is a simple consequence
of the following bound: consider a baseball league that consists of teams
with either strengths $x=1$ or $x=x_{\rm min}>0$, and with games only between
strong and weak teams.  Then the distribution of winning streaks of the
strong teams decays as $(1+x_{\rm min})^{-n}$; this represents an obvious
upper bound for the streak distribution in a league where team strengths are
uniformly distributed in $[x_{\rm min},1]$.

We now apply the BT model to determine the form of the consecutive-game
winning and losing streak distributions.  Using Eq.~\eqref{W} for the
single-game outcome probability, the probability that a team of strength $x$
has a streak of $n$ consecutive wins is
\begin{equation}
\label{Wn}
P_n(x) = \prod_{j=1}^n \frac{x}{x+x_j}\,\,\, \frac{x_{0}}{x+x_{n+1}}\,\,\,
\frac{x_{n+1}}{x+x_{n+1}}~.
\end{equation}
The product gives the probability for $n$ consecutive wins against teams of
strengths $x_j$, $j=1,2,\ldots,n$ (some factors possibly repeated), while the
last two factors give the probability that the $0^{\rm th}$ and the
$(n+1)^{\rm st}$ games are losses to terminate the winning streak at $n$
games.  Assuming a uniform team strength distribution $\rho(x)$, and for the
case where each team plays the same number of games with every opponent, we
average Eq.~\eqref{Wn} over all opponents and then over all teams.

The first
average gives:
\begin{equation}
\label{Wn-fac}
\langle P_n(x)\rangle_{\{x_j\}} = {x}^n\left\langle
  \frac{1}{x+y}\right\rangle^n \left\langle
  \frac{y}{x+y}\right\rangle~.
\end{equation}
with
%\begin{subequations}
%\label{avs}
\begin{align}
\left\langle\frac{1}{x+y}\right\rangle
%&= \frac{1}{1-\epsilon}\int_{\epsilon}^{1} \frac{dy}{x+y}\nonumber \\
 &= \frac{1}{1-\epsilon}\ln\left(\frac{x+1}{x+\epsilon}\right)~,\nonumber\\
\left\langle\frac{y}{x+y}\right\rangle
 &= 1- \frac{x}{1-\epsilon}\ln\left(\frac{x+1}{x+\epsilon}\right)~\nonumber
\end{align}
%\end{subequations}
for a uniform distribution of team strengths in $[x_{\rm min},1]$.  Here we
use the fact that each team strength is independent, so that the product in
Eq.~\eqref{Wn} factorizes.
We now average over the uniform strength distribution, to find, for the
team-averaged probability to have a streak of $n$ consecutive wins,
\begin{eqnarray}
\label{av2Wn}
\langle P_n\rangle
= \frac{1}{1-x_{\rm min}}\int_{x_{\rm min}}^{1} f(x)\,\, e^{n g(x)}\, dx\,,
\end{eqnarray}
where
\begin{eqnarray*}
f(x)&=&\left[1
  -\frac{x}{1-x_{\rm min}}\ln\left(\frac{x+1}{x+x_{\rm min}}\right)\right]^2\\
g(x)&=&\ln x + \ln\left[\frac{1}{1-x_{\rm min}}
  \ln\left(\frac{x+1}{x+x_{\rm min}}\right)\right].
\end{eqnarray*}
Since $g(x)$ monotonically increases with $x$ within $[x_{\rm min},1]$, the
integral in Eq.~\eqref{av2Wn} is dominated by the behavior near the maximum
of $g(x)$ at $x=1$ for large $n$.  Performing the integral by parts
\cite{BO}, the leading behavior is
\begin{equation}
\label{Wav}
\langle P_n\rangle \sim e^{n g(1)}~,
\end{equation}
with
\begin{equation*}
\qquad g(1)=-\ln(1-x_{\rm min}) + \ln \ln\left(\frac{2}{1+x_{\rm min}}\right)~.
\end{equation*}

As expected, $\langle P_n\rangle$ decays exponentially with $n$, but with a
decay rate that decreases as teams become more heterogeneous (decreasing
$x_{\rm min}$).  In the limit of equal-strength teams, the most rapid decay
of the streak probability arises, $P_n=2^{-n}$, while the widest disparity in
team strengths, $x_{\rm min}=0$, leads to the slowest possible decay $P_n\sim
(\ln 2)^n\approx (0.693)^n$.

We simulated the streak distribution $P_n$ using the same methodology as that
for the win/loss records; related simulations of streak statistics are given
in Refs.~\cite{JAS93,A04}.  Taking $x_{\rm min}=0.435$ for 1961--2005---the
same value as those used in simulations of the win/loss records---we find a
good match to the streak data for this period.  The apparent systematic
discrepancy between data and theory for $n\geq 17$ is illusory because
streaks do not exist for every value of $n$.  Moreover, the number of streaks
of length $n\geq 17$ is only eight, so that fluctuations are quite important.

For the 1901--60 period, if we use $x_{\rm min}=0.278$, the data for $P_n$ is
in excellent agreement with theory for $n< 17$.  However, for $n$ in the
range 17--22, the data is a roughly factor of 2 greater than that given by
the analytical solution Eq.~\eqref{Wav} or by simulations of the BT model.
Thus the tail of the streak distribution for this early period appears to
disagree with a purely statistical model of streaks.  Again, the number of
events for a $n\geq 17$ is 5 or less, compared to a total number of $\sim
70000$ winning and losing streaks during this period.  Hence one cannot
exclude the possibility that the observed discrepancy for $n\geq 17$ is
simply due to lack of statistics.

Finally, we test for the possible role of self-reinforcement on winning and
losing streaks.  To this end, we take each of the 2166 season-by-season
win/loss histories for each team and randomize them $10^5$ times.  For each
such realization of a randomized history, we compute the streak distribution
and superpose the results for all randomized histories.  The large amount of
data gives streak distributions with negligible fluctuations up to $n=30$ and
which extend to $n=44$ and 41 for the two successive periods.  More
strikingly, these streak distributions based on randomized win/loss records
are virtually identical to the simulated streak data as well as to the
numerical integration of Eq.~\eqref{av2Wn}, as shown in
Fig.~\ref{streaks-sep}.

\section{Summary}

To conclude, the Bradley-Terry (BT) competition model, in which the outcome
of any game depends only on the relative strengths of the two competing
teams, quantitatively accounts for the average win/loss records of
Major-League baseball teams.  The distribution of team strengths that gives
the best match to these win/loss records was found to be quite close to
uniform over a range $[x_{\rm min},1]$, with $x_{\rm min}\approx 0.28$ for
the early modern era of 1901--1960 and $x_{\rm min}\approx 0.44$ for the
expansion era of 1961--2005.  This same BT model also reproduces the
season-to-season fluctuations of the win/loss records.  An important
consequence of the BT model is the existence of a non trivial
detailed-balance relation which we verified with satisfying accuracy.  We
consider this verification as a quite stringent test of the theory.

The same BT model was also used to account for the distribution of team
consecutive-game winning and losing streaks.  We found excellent agreement
between the prediction of the BT model and the streak data for $n<17$ for
both the 1901-60 and 1961-2005 periods.  However, the tail of the streak
distribution for the 1901--60 period with $n\geq 17$ is less accurately
described by the BT theory and it is an open question about the mechanisms
for the discrepancy, although it could well originate from lack of
statistics.  We also provided evidence that self-reinforcement plays little
role in streaks, as randomizations of the actual win/loss records produces
streak distributions that are indistinguishable from the streak data except
in for the $n\geq 17$ tail during the 1901-60 period.

We also showed that the optimal team strength distribution is narrower for
the period 1961--2005 compared to 1901--60.  This narrowing shows that
baseball competition is becoming keener so that outliers in team performance
over an entire season---as quantified by win/loss records and lengths of
winning and losing streaks---are less likely to occur.

We close by emphasizing the parsimonious nature of our modeling. The only
assumed features are the Bradley-Terry form Eq.~\eqref{W} for the outcome of
a single game, and the uniform distribution of the winning probabilities,
controlled by the single free parameter $x_{\rm min}$.  All other model
features can then be inferred from the data.  While we have ignored many
aspects of baseball that ought to play some role---the strength of a team
changing during a season due to major trades of players and/or injuries,
home-field advantage, etc.---the agreement between the win fraction data and
the streak data with the predictions of the Bradley-Terry model are extremely
good.  It will be worthwhile to apply the approaches of this paper to other
major sports to learn about possible universalities and idiosyncracies in the
statistical features of game outcomes.

{\bf Acknowledgments:} SR thanks Guoan Hu for data collection assistance, Jim
Albert for literature advice, and financial support from NSF grant DMR0535503
and Universit\'e Paul Sabatier.

\newpage
\appendix*
\section{Team Winning and Losing Streaks}

\begin{minipage}{3.5in}

{\small\begin{longtable}{|p{0.2in}|p{0.3in}|p{1.8in}|}
\caption{Winning streaks of $n\geq 15$ games since 1901.}\label{w-streak}\\
\hline
n   &year   &team  \\ \hline

26&   1916& New York Giants (1 tie)\\ \hline
21&   1935& Chicago Cubs\\ \hline
20&   2002& Oakland Athletics\\ \hline
19&   1906& Chicago White Sox (1 tie)\\ \hline
19&   1947& New York Yankees\\ \hline
18&   1904& New York Giants\\ \hline
18&   1953& New York Yankees\\ \hline
17&   1907& New York Giants\\ \hline
17&   1912& Washington Senators\\ \hline
17&   1916& New York Giants\\ \hline
17&   1931& Philadelphia Athletics\\ \hline
16&   1909& Pittsburgh Pirates\\ \hline
16&   1912& New York Giants\\ \hline
16&   1926& New York Yankees\\ \hline
16&   1951& New York Giants\\ \hline
16&   1977& Kansas City Royals\\ \hline
15&   1903& Pittsburgh Pirates\\ \hline
15&   1906& New York Highlanders\\ \hline
15&   1913& Philadelphia Athletics\\ \hline
15&   1924& Brooklyn Dodgers\\ \hline
15&   1936& Chicago Cubs\\ \hline
15&   1936& New York Giants\\ \hline
15&   1946& Boston Red Sox\\ \hline
15&   1960& New York Yankees\\ \hline
15&   1991& Minnesota Twins\\ \hline
15&   2000& Atlanta Braves\\ \hline
15&   2001& Seattle Mariners \\ \hline
\end{longtable}
}

 \end{minipage}
  \begin{minipage}{3.5in}

{\small\begin{longtable}{|p{0.2in}|p{0.3in}|p{2.2in}|}
\caption{Losing streaks of $n\geq 15$ games since 1901.}\label{l-streak}\\
\hline
n  &year   &team  \\ \hline

23&   1961& Philadelphia Phillies\\ \hline
21&   1988& Baltimore Orioles\\ \hline
20&   1906& Boston Americans\\ \hline
20&   1906& Philadelphia As\\ \hline
20&   1916& Philadelphia As\\ \hline
20&   1969& Montreal Expos     (first year)\\ \hline
19&   1906& Boston Beaneaters\\ \hline
19&   1914& Cincinnati Reds\\ \hline
19&   1975& Detroit Tigers         \\ \hline
19&   2005& Kansas City Royals\\ \hline
18&   1920& Philadelphia As       \\ \hline
18&   1948& Washington Senators    \\ \hline
18&   1959& Washington Senators    \\ \hline
17&   1926& Boston Red Sox     \\ \hline
17&   1962& NY Mets            (first year)\\ \hline
17&   1977& Atlanta Braves\\ \hline
16&   1911& Boston Braves\\ \hline
16&   1907& Boston Doves\\ \hline
16&   1907& Boston Americans   (2 ties)\\ \hline
16&   1944& Brooklyn Dodgers   (1 made-up game)\\ \hline
15&   1909& St. Louis Browns\\ \hline
15&   1911& Boston Rustlers    \\ \hline
15&   1927& Boston Braves      \\ \hline
15&   1927& Boston Red Sox     \\ \hline
15&   1935& Boston Braves\\ \hline
15&   1937& Philadelphia As        \\ \hline
15&   2002& Tampa Bay          \\ \hline
15&   1972& Texas Rangers      (first year)\\ \hline
\end{longtable}
}
  \end{minipage}

%\end{widetext}


\begin{thebibliography}{99}
\bibitem{contrast} In contrast, in Ref.~\cite{BVR}, the winning probability
  was taken to be independent of the relative strengths of the two teams; the
  stronger team won with a fixed probability $p$ and the weaker won with
  probability $1-p$.
\end{thebibliography}
\end{document}